# Automatic Segmentation of Vertebral Features on Ultrasound Spine Images using Stacked Hourglass Network

Hong-Ye Zeng, Song-Han Ge, Yu-Chong Gao, De-Sen Zhou, Kang Zhou, Xu-Ming He, Edmond Lou, Rui Zheng*, *Member, IEEE*

*Abstract*— *Objective:* The spinous process angle (SPA) is one of the essential parameters to denote three-dimensional (3-D) deformity of spine. We propose an automatic segmentation method based on Stacked Hourglass Network (SHN) to detect the spinous processes (SP) on ultrasound (US) spine images and to measure the SPAs of clinical scoliotic subjects. *Methods:* The network was trained to detect vertebral SP and laminae as landmarks on 1200 ultrasound transverse images and validated on 100 images. All the processed transverse images with highlighted SP and laminae were reconstructed into a 3D image volume, and the SPAs were measured on the projected coronal images. The trained network was tested on 400 images by calculating the percentage of correct keypoints (PCK); and the SPA measurements were evaluated on 50 scoliotic subjects by comparing the results from US images and radiographs. *Results:* The trained network achieved a high average PCK (86.8%) on the test datasets, particularly the PCK of SP detection was 90.3%. The SPAs measured from US and radiographic methods showed good correlation (r>0.85), and the mean absolute differences (MAD) between two modalities were 3.3°, which was less than the clinical acceptance error (5°). *Conclusion:* The vertebral features can be accurately segmented on US spine images using SHN, and the measurement results of SPA from US data was comparable to the gold standard from radiography.

## I. Introduction

Adolescent idiopathic scoliosis (AIS) is a 3D deformity of spine featured with lateral deviation and axial vertebral rotation [1] [2]. In China, the prevalence rate of AIS is approximate to 5.14%, especially the high prevalence of 13.81% occurs in girls aged 14-15 [3]. AIS can greatly influence the patient's life due to back pain and obvious appearance changes, and large curvature can cause distorted rib cage and bring pressure for lung and heart, even give rise to death. Therefore, the early diagnosis and frequent follow-up exams of AIS are very important to the treatment. The Cobb angle measured on posterior-anterior radiographs is the gold standard for diagnosing and monitoring spinal curvature, and commonly used for treatment options and assessment of treatment outcome [4,5,6]. The spinous process angle (SPA) is an alternative measurement parameter to define the posterior deformity of spine, which can reveal the information of both the axial vertebral rotation and lateral curvatures [7].

The manual measurement of scoliotic curves depended on the trained specialists, who were required to identify the vertebrae using measurement tools. However, it could take plenty of time for the manual measurement even for an experienced rater. The automatic measurement of Cobb angle on radiographs has been proposed [8-12]. Horng et al. [8] applied the U-Net, Dense U-Net and Residual U-Net to segment the vertebral body on radiographs, and found that the Residual U-Net was superior to other two networks. Afterwards, the segmentation results were used to measure the Cobb angle by using minimum bounding rectangle method, and there showed no significant differences between automatic and manual measurement. One vertebra could also be described as four landmarks, which were located at the four vertices of vertebra. Wu et al. [9] proposed the BoostNet to detect the landmarks robustly and accurately. The BoostLayer was used to eliminate the outlier features, and the spinal structural multi-output layer was used to improve detection accuracy by utilizing the prior knowledge of structure. Sun et al. [10] proposed Support Vector Regression (S$^2$VR) method to predict the landmarks and estimate Cobb angles directly from radiographic features, and the results showed great effectiveness comparing to baseline methods. Zhang et al. [11] applied the HR-Net to detect the landmarks and output Cobb angle on non-original radiographs acquired by smartphone. The predictions from network were still accurate under the circumstance of low image qualities. Wu et al. [12] applied the multi-view correlation network (MVC-Net) to find spinal landmarks and estimate Cobb angle on anterior-posterior (AP) and lateral (LAT) radiographs, and had achieved robust and

This work was sponsored by Natural Science Foundation of Shanghai under Grant No.19ZR1433800. (Corresponding author: Rui Zheng)

Hong-Ye Zeng, Song-Han Ge, Yu-Chong Gao, De-Sen Zhou and Kang Zhou are with School of Information Science and Technology, ShanghaiTech University, Shanghai 201210, China, Shanghai Institute of Microsystem and Information Technology, Chinese Academy of Sciences, Shanghai 200050, China, and University of Chinese Academy of Sciences, Beijing 100049, China. (e-mail: zenghy@shanghaitech.edu.cn, gesonghan@shanghaitech.edu.cn, gaoych@shanghaitech.edu.cn, zhouds@shanghaitech.edu.cn, zhoukang@shanghaitech.edu.cn).

Edmond Lou, is with Faculty of Engineering - Electrical & Computer Engineering Dept., University of Alberta, 8440 -112, Edmonton T6G 2B7, Canada. (e-mail: elou@ualberta.ca)

Rui Zheng, Xu-Ming He is with School of Information Science and Technology and Shanghai Engineering Research Center of Intelligent Vision and Imaging, ShanghaiTech University, Shanghai 201210, China (e-mail: zhengrui@shanghaitech.edu.cn, hexm@shanghaitech.edu.cn).

accurate result on 526 x-ray images.

The radiography has been widely used for Cobb angle measurements, but it can accumulate ionizing radiation for patients during the long-term follow-up exams, furthermore cause high risk of cancer especially for females [13,14,15,16]. Compared to radiography, the ultrasound (US) imaging technique is radiation-free, cost-effective and portable. Recently, the 3-D US imaging techniques with application of the center of lamina (COL) [17] method, the aid of previous radiograph (AOR) [18] and the volume projection imaging (VPI) method [19] have been introduced to directly measure the spinal curvatures. The intra- and inter-rater reliabilities of the US method for clinical cases were validated [20], and the factors influencing spine curve measurement were explored [21]. The automatic measurement of spinal curvature was implemented on the VPI images by detecting spinous column profile and extracting spine curves, and the results demonstrated a good correlation to the manual results ($r = 0.90$, $p < 0.001$) and to the radiographic Cobb's angles ($r = 0.83$, $p < 0.001$) [22].

Even though the reliable assessment of scoliotic curves has been achieved from ultrasound spine images, current measurements are still focusing on the proxy Cobb angles and lack of SPA information. The spinous process is usually merged into soft tissue and cannot be directly marked on coronal images, meanwhile it is impossible to manually specify the SP on hundreds of transverse frames. Thus, the SPA estimation is a big challenge for the diagnosis using ultrasound imaging methods. To implement SPA measurement, the automatic method for vertebral feature identification is an essential step of finding spinous process curves.

The convolutional neural network has been considered as one solution for bone identification on ultrasound images. Wang et al. [23] used a pre-enhancing net to enhance the bone surface, and then applied the modified U-net with classification layer to segment and classify the knee, femur, distal, radius and tibia bones. Alsinan et al. [24] proposed a network to segment the radius, femur and tibia bones, which fused the B-mode US image with local phase features. The method obtained an average F-score above 95% and average bone surface localization error of 0.2 mm, which were superior to other relative methods. Ungi et al. [25] applied the U-Net to segment and visualize the transverse process and ribs; the segmented result was reconstructed to a spine volume and could be used to measure the spinal curvature.

In our previous study, the GVF snake model has been applied to detect the spinous process tips on transverse images, and the manual and automatic results of SPA assessment were obtained from US scans [26,27]. However, the GVF method was very sensitive to image quality, and the whole process was time consuming and impractical for clinical application. In this study, we apply the stacked hourglass network to detect the SP and laminae position as landmarks on transverse ultrasound images. The stacked hourglass network (SHN) [28] is a commonly used convolutional neural network to estimate human pose, which can process features across all scale to capture the coherent understanding of the full image. Like the human joints, there is strong correlation between SP and laminae according to the vertebral anatomical structures. Thus, the SHN has the potential to identify the SP and laminae as landmarks.

This paper is organized as follow. Section II introduces the procedures of data collection and the network architecture. The evaluation metrics for trained network and the assessment for SPA on the reconstructed coronal images are also included. Section III shows the results of SP detection and the SPA measurement based on the processed images. The comparison with GVF method, drawbacks of proposed method and future works are discussed in Section IV, and Section VI concludes the proposed method in this study.

## II. MATERIAL AND METHODS

### A. Subjects and Data Acquisition

The ultrasound spine data were collected from 92 adolescents (17 males and 75 females; mean age:14.3±1.9; body mass index: 14.3~30.8). They fulfilled the following inclusion criteria: 1) no prior surgical treatment; 2) received a posterior-anterior standing radiograph and US scan within an hour; 3) with the major curve of 10° to 45°. Before being enrolled into the study, all participating subjects signed a written consent, and ethics approval was granted from the local health research ethics board. The radiographic curve information including SPAs and Cobb angles were exported from the medical record to be used for the following data analysis.

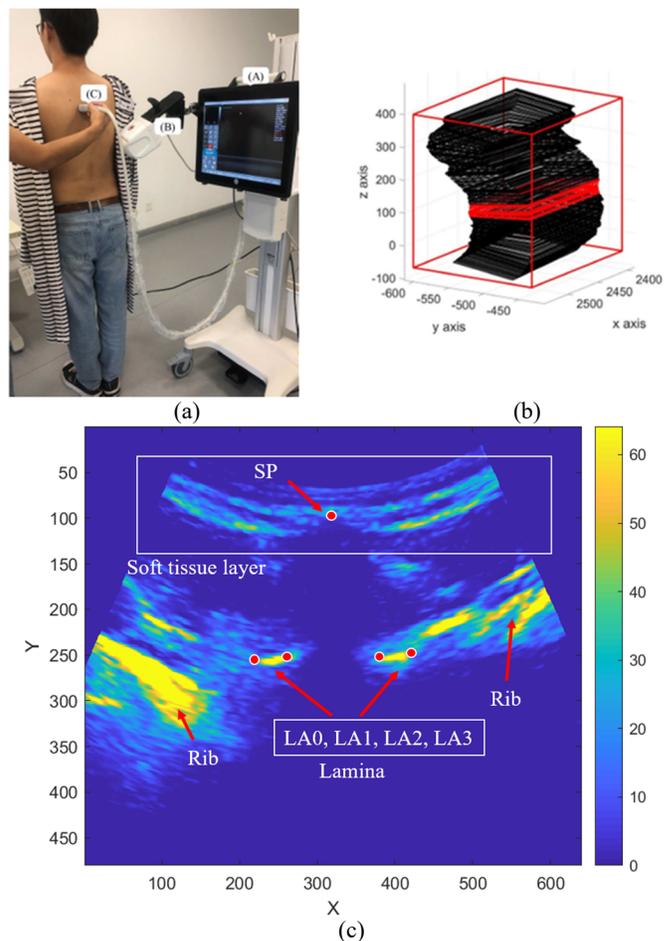

Fig. 1. Data acquisition: (a) the ultrasound image acquisition system; (b) the demonstration of a series of acquired 2D transverse frames from one data set; (c) one transverse ultrasound image. The 'SP' point represented spinous process; four points 'LA0, LA1, LA2, LA3' represented the proxy endpoints of laminae.

The standing ultrasound scan along full spine was obtained using the SonixONE system equipped with 128-element C5-2/60 array transducer and SonixGPS (Analogic Ultrasound - BK Medical, Peabody, Massachusetts, US) as shown in Fig. 1a. The center frequency of transducer probe was 2.5 MHz, and the imaging depth was set as 6cm.

One ultrasound spine scan contained approximate 900-2300 transverse frames acquired along the spine curve illustrated as Fig. 1b, and Fig. 1c showed a typical transvers image frame with vertebral features. In this study, forty-two ultrasound scans were selected as the source of the training and testing dataset, and fifty ultrasound scans were used to validate the SPA assessment.

*B. Ground-Truth Labelling*

The acquired transverse frames were 8-bit 640*480 grayscale images. The spinous process and laminae endpoints were designated as five landmarks and illustrated as red points in Fig. 1c. The labelling was implemented by two raters, who had been trained on over 100 US images with typical vertebral structures in order to accurately identify the features to be designated.

For the 42 ultrasound scans which were selected to extract the training and testing data, fifty transverse images were collected randomly from each scan and labeled for the 5 landmarks by raters. All the frames were also screened and would be abandoned if it was annotated with large difference between two raters, or was identified as deficient structure with unclear laminae and SP. In the end, the entire dataset contained 1700 ultrasound transverse images, in which 1200 were used for training, 100 for validation and 400 for testing, respectively.

*C. Landmark Detection Network*

The proposed image processing procedure was illustrated in Fig. 2, and two stacked hourglasses blocks were implemented in the experiments. To avoid the demand of a significant amount of GPU memory, the transverse frame (640*480) from original ultrasound image sequence was resized to 256*256, and the stride 2 convolutional and max pooling layer were used to further reduce the resolution to 64*64. The convolutional layer simultaneously extracted the feature map (64*64*256) from image, which would be input into the hourglass.

The architecture of single hourglass block was illustrated in Fig. 3. In the hourglass, the network applied the max pooling layer 4 times to reduce the resolution of feature map from 64*64 to 4*4, and used the residual block after each downsampling to capture spatial context. Meanwhile, the skip layer was used to preserve the spatial information which was lost in the max pooling layers at each resolution. After reaching the lowest resolution (4*4), the network began to restore the resolution by applying the nearest neighbor upsampling, and then to recover spatial information by adding the feature map preserved in skip layer. The channel number of feature maps maintained 256 across all scales, and the processed feature maps could be used to extract the predicted heatmaps $H_0$.

The single hourglass was applied to capture global and local features, and the stacked multiple hourglasses could reprocess the features to capture higher-level information. The input and output of first hourglass and heatmaps $H_0$ were combined as the input of the second hourglass. Through the stacked hourglass network, two sets of heatmaps $H_0$ and $H_1$ were finally obtained, and the loss was applied onto all heatmaps to implement the intermediate supervision, only $H_1$ produced the final predictions for the location of landmarks.

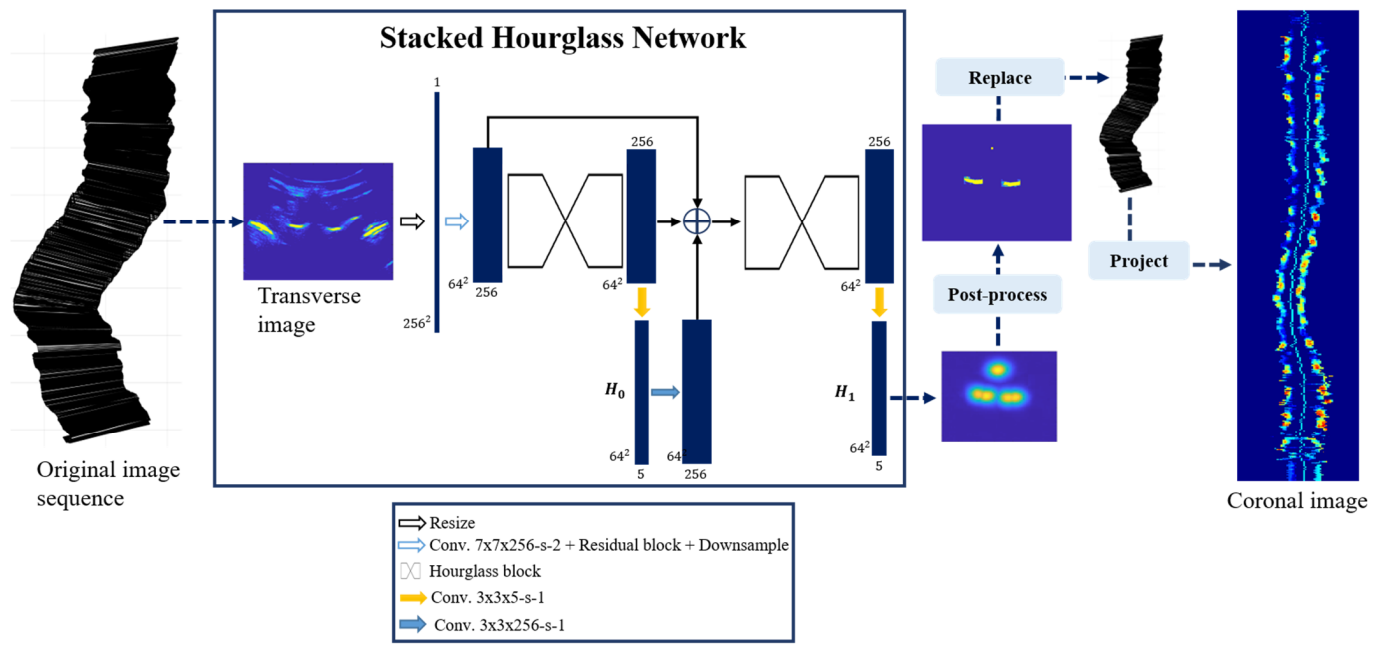

Fig. 2. Overview of the proposed method. The transverse image in ultrasound data sequence was firstly resized, reduced to the specific resolution, and then input into the stacked hourglass network. The SHN consisted of two hourglasses and output the predicted heatmaps. The positions of SP and laminae were predicted from the heatmaps $H_1$. The transverse image was processed according to the prediction and replaced the original frame. Afterwards, the US scan was reconstructed and the coronal image was obtained. In the figure, each solid blue box is an image or multi-channel feature map. Each arrow corresponds to the detailed operations as shown in the legend. The number of channels was denoted on the top and bottom of blue box. The resolution was provided at the lower left edge of the blue box.

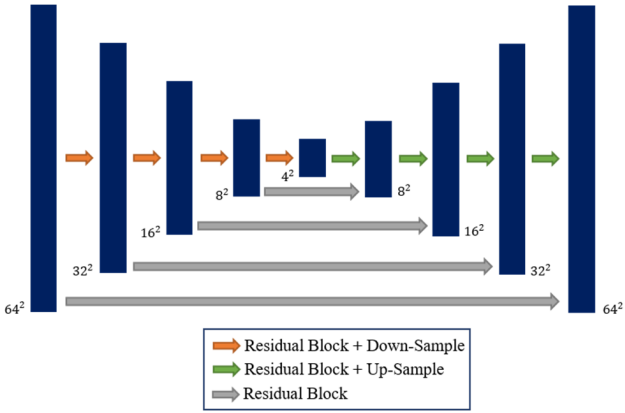

Fig. 3. Demonstration of a single hourglass block.

### D. Training Details

The network was trained from scratch using Pytorch, and the training process took 1000 epochs (about four hours) on two 8 GB NVIDIA RTX 2070 GPUs. To avoid overfitting, the data augmentation was implemented including random rotation (+/- 20 degree) and random flip around vertical axis (p=0.5). The logarithm transformation was used to enhance the soft tissue layer, since the low intensity would cause low SNR of image and result in lower detection accuracy of spinous process landmark.

The ground-truth heatmaps $H$ was a sets of 2-D Gaussian distribution centered at the labeled landmarks with standard deviation of 4 pixels. The loss was calculated between ground-truth heatmaps and two sets of predicted heatmaps ($H_0$ and $H_1$) by using mean square error (MSE). The Adam was used to optimize the training loss with learning rate of 1e-5 on 500 epochs, and 1e-7 on 500 epochs.

### E. Position Extraction and Post-process

The network output five 64*64 heatmaps $H_1$ which were corresponding to five landmarks representing SP and four lamina endpoints. Given a heatmap $h$ in $H_1$, the coordinates of maximal and second maximal activation were found as $\overrightarrow{m_1}$ and $\overrightarrow{m_2}$, and the coordinate of the predicted landmark $\vec{p}$ was identified as [28]:

$$\vec{p} = \overrightarrow{m_1} + 0.25 \frac{\overrightarrow{m_2} - \overrightarrow{m_1}}{\|\overrightarrow{m_2} - \overrightarrow{m_1}\|} \quad (1)$$

where $\|\cdot\|$ was the magnitude of vector. Afterwards, the predicted coordinate $\vec{p}$ on 64*64 heatmap was converted to final coordinate $\overrightarrow{p'}$ on original 640*480 image as follow:

$$\overrightarrow{p'} = [\gamma_1 \quad \gamma_2] * \vec{p} \quad (2)$$

where $\gamma_1 = 640/64$ and $\gamma_2 = 480/64$ were resolution reduction ratios.

The predicted landmarks were then verified as following criterions: 1) five landmarks were arranged in the order of left lamina (LA0 and LA1), SP, right lamina (LA2 and LA3) in x-axis direction as shown in Fig 1c; 2) the distance between two landmarks from the same lamina (e.g. LA0&LA1 or LA2&LA3) was within the range of 10-80 pixels. To highlight SP for further projection to coronal plane, the coordinates of the detected SP and its 8-neighbor pixels was assigned to 255. A rectangular region around the detected lamina positions was retained to remove ribs and soft tissue layer.

### F. SPA Measurement

After landmark position extraction and post-processing, the processed transverse frames replaced the original frames, and Voxel-Based Nearest Neighbor (VNN) and Fast Dot Projection (FDP) method [29] were used to reconstruct the 3-D data volume and to obtain coronal image of spine as show in Fig 2. The spinous process curve was indicated by fitting all the identified SP points on the coronal image into a 5$^{th}$ polynomial curve, and the SPA was measured as proposed in our previous work [27].

### G. Evaluation Metrics

The evaluation of landmark detection was based on the standard Percentage of Correct Keypoints (PCK) metric, which reported the percentage of detections that were within the region of the ground truth. In this study, the acceptable region between prediction and ground truth was a circle with radius of 15 pixels under the circumstance that the original image resolution was 480*640.

The method was also validated by calculating the reliability and accuracy of the SPA measurement comparing to the gold standard from radiographs. The intra-class correlation coefficients (ICC [2,1]) was computed to evaluate the reliability between different raters. The mean absolute difference (MAD), standard deviation (SD) and correlation coefficient between US and radiographic measurements were used to assess the accuracy and robustness of the method.

## III. RESULT

Table I listed the detection results of different landmarks for the test data set which included 400 transverse ultrasound images. The trained network showed high PCK on detecting the position of spinous process and laminae, especially the percentage of the correctly detected SP was 90.3%. Fig. 4 showed some predictions of various transverse images with different rotation angles and different status of lamina pairs.

Fifty ultrasound scans were applied for the SPA measurement. It averagely took about ten minutes to detected SP and laminae positions using stacked hourglass network for each scan. After the 3-D reconstruction, the SP and laminae were highlighted on the ultrasound coronal image as shown in Fig. 5a. Seventy-six SP curves were attained from the 50 data sets and manually measured by two raters.

Table II showed the inter-rater measurement difference when using ultrasound method, the ICC value manifested high reliability between different raters. Fig. 5b&c illustrated an example of SPA measurement by two modalities. As shown in the white box, the SP curves revealed apparent vertebral rotation and were deviated to the same direction (from the middle line of vertebra to the right side) on both ultrasound image and radiograph. It indicated that the SP curves obtained by US qualitatively kept the similar trend as radiography. Table III illustrated high agreement of the SPA measurements using two methods quantitatively, and the average measurement differences were less than the clinical acceptance error (5°).

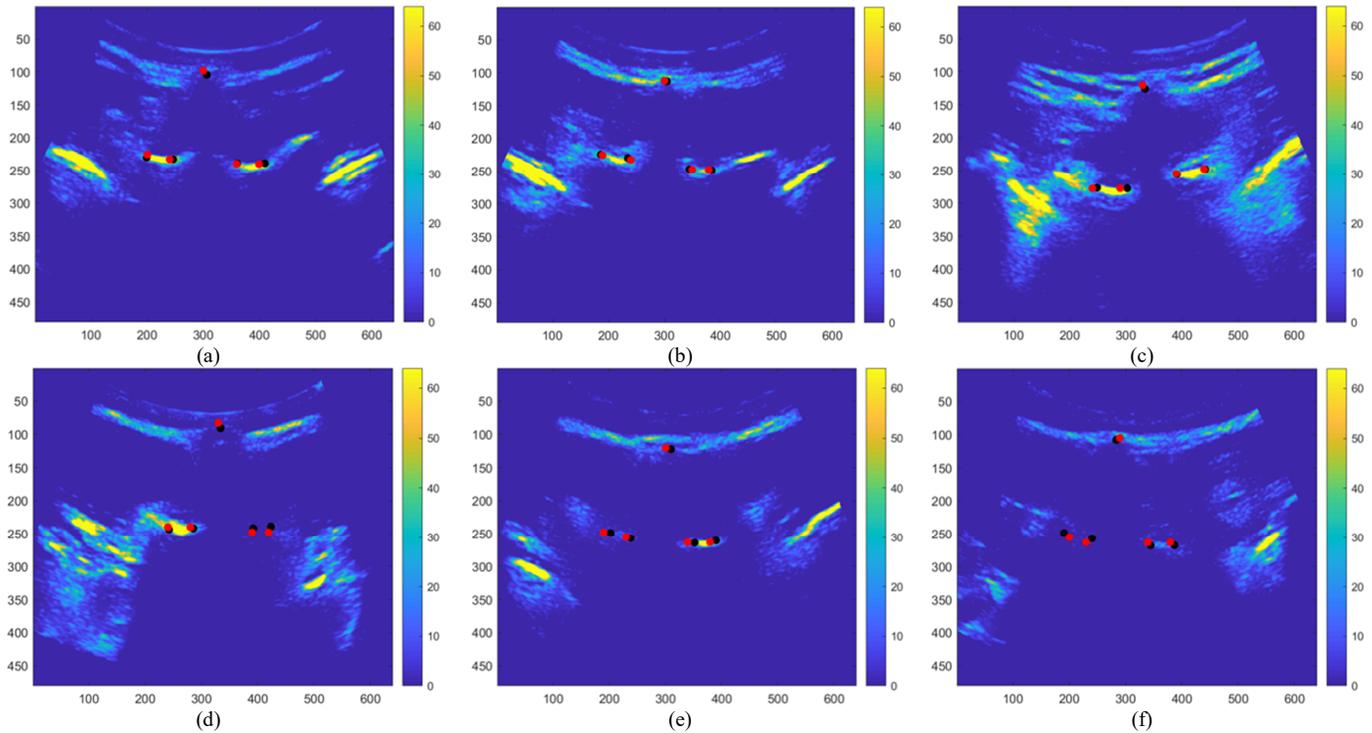

Fig. 4. Some accurately predicted keypoints on various transverse ultrasound images from validation dataset. Black points were ground truth, and red points were predictions. (a)&(b)&(c) contained two clear laminae but rotated to different angles, (d)&(e)&(f) only contained one clear lamina and the other lamina was invisible.

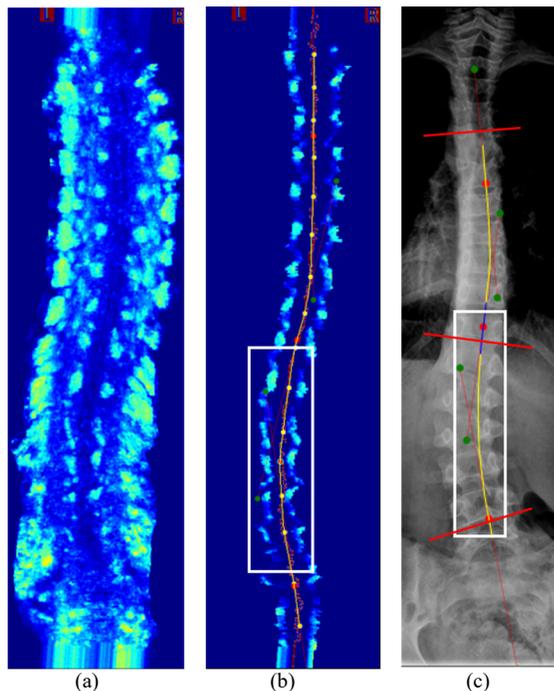

Fig. 5. The demonstration of SPA detection and measurement: (a) original ultrasound coronal image; (b) the processed coronal ultrasound image with SP and laminae; and (c) the SPA measurement on radiograph. The SP curves were obtained by fitting all detected SP points to a $5^{th}$ order polynomial. The SPAs were 14/23° on (b) and 14/20° on (c). The Cobb angle on (c) were 22/27°.

Fig. 6a&b presented the comparison of SPA between two modalities, and the correlations were both higher than 0.85 for two raters. There were total 16 curves showed large differences (>5°) among all measurements from two raters, and the mean of these differences were 8.0±2.7°, 7.9±1.6° for Rater 1 and 2 respectively.

Fig. 7a&b illustrated the comparison between SPAs from ultrasound measurements and Cobb angles from radiographs. The correlations between two curvatures were 0.75 and 0.80, and the MAD were 5.7±4.5° and 6.1±4.8° for two raters, respectively. There were 41 curves with difference larger than 5°, and 31 of them illustrated that the Cobb angles far exceeded the SPAs.

IV. DISCUSSION

Our stacked hourglass network consisted of two hourglasses. Four or eight stacked hourglasses can achieve modest improvement according to Newell's investigation [30], but it caused overfitting problem due to excessive network parameters in our study. The structure of two hourglasses could avoid overfitting and still output accurate predictions.

In addition to the proposed method for position extraction from heatmaps, we also tried the distribution-aware coordinate representation of keypoints (DARK) method [30], which modulated the predicted heatmap distribution before extracting position. Only 0.1% improvement of total landmark detection accuracy was acquired, and the SPA measurement accuracy remained the same. It indicated that our output heatmaps basically followed the 2-D Gaussian distribution, and the DARK method was not essential in our method.

TABLE I PERCENTAGE OF THE CORRECTLY DETECTED KEYPOINTS FOR TEST DATASET.

| SP | LA0 | LA1 | LA2 | LA3 | Total |
|---|---|---|---|---|---|
| 90.3 | 84.3 | 88.5 | 87.0 | 83.7 | 86.8 |

TABLE II THE INTER-RATER ANALYSIS OF ULTRASOUND SPA MEASUREMENT.

| Method | Rater | MAD(°) | SD (°) | ICC (2,1) |
|---|---|---|---|---|
| US | R1 versus R2 | 2.3 | 2.0 | 0.93 |

TABLE III THE COMPARISON OF SPA MEASUREMENT BETWEEN THE PROPOSED ULTRASOUND METHOD AND RADIOGRAPHY.

| Rater | MAD(°) | SD (°) | Max difference(°) | Number of measurement differences larger than 6° | Correlation coefficient |
|---|---|---|---|---|---|
| R1 | 3.3 | 2.6 | 14 | 14 | 0.89 |
| R2 | 3.3 | 2.3 | 12 | 10 | 0.87 |

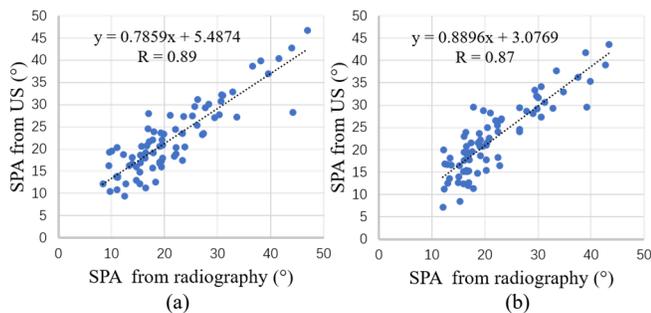

Fig. 6. The comparison between ultrasound and radiographic measurements of SPAs from a) rater 1 and b) rater 2.

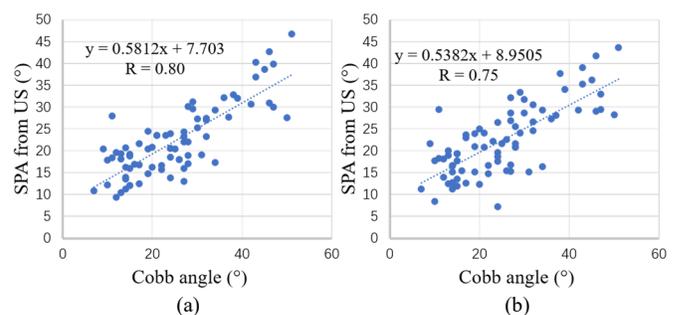

Fig. 7. The comparison between SPAs and Cobb angles from a) rater 1 and b) rater 2.

The stacked hourglass network was trained for SP and laminae detection, and acquired good performance on the test dataset. Commonly, the laminae are easy to detect, because they are usually located in the middle region of the transverse image with high intensity features. Meanwhile, there is strong correlation between spinous process and laminae due to anatomical structures of vertebra, which can be utilized by network and make the prediction easier and better. Based on this hypothesis, the laminae were annotated to help predict position of spinous process, and furthermore the detected laminae could be used to outline the structure of posterior vertebrae. However, as illustrated in Table I, the detection accuracy of SP was slightly higher than laminae. Although the SHN detected the laminae accurately and showed good inferencing ability of invisible laminae as shown in Fig 4d-f in some cases, the complex bone and soft tissue structures on transverse images still challenged the network. Fig 8 illustrated three main factors disturbing the prediction of landmarks, especially for laminae. Firstly, the inferior articular processes from the upper vertebra and the lamina from the lower vertebra could appear on the same transvers image since these two structures were anatomically overlapped each other. In Fig. 8a, the inferior articular process with high intensity from upper vertebra (indicated in the white box) was identified as the right lamina candidate by network and shown in the heatmaps together with correct prediction of the lamina. Therefore, there were two major activations on the heatmap of LA2 and LA3. Secondly, the multiple soft-tissue layers in the lower back could produce high intensity features caused by the reflection from the interface between different tissues. As shown in the lumbar vertebra frame (Fig. 8b), even though the confident predictions of left lamina (LA0&LA1) were obtained and the intensity of right lamina (LA2&LA3) was high, the white box region with similar intensity was still recognized as laminae candidate as the heatmaps displayed. Thirdly, in Fig. 8c, there were three high intensity areas equally spaced in the middle of image, but the rightmost area indicated rib according to the anatomical symmetry of vertebra. As the result, the network output wrong position of laminae due to the similar features, and consequently the SP was predicted mistakenly since it was required to be in the middle of two "identified laminae". In summary, the multiple high intensity features near lamina region caused by complex structures could result in multiple centers on heatmaps and influence the predictions. Therefore, to improve the detection accuracy, more data should be supplemented to cover different situations in the future.

The reliable SPA results were acquired on segmented coronal images, but discrepant measurements occurred occasionally due to the following factors. Firstly, the different standing postures might induce the change of spinal curvature on ultrasound images and radiographs, particularly it mostly happened on the cases with mild curves (<25°), as shown in Fig. 9a-c. Secondly, the poor image quality of the ultrasound frames acquired from lumbar area would cause wrong SP detection results, and furthermore produce scattered SP positions on the coronal images as shown in the red box of Fig. 10a. Thirdly, during the scan, the probe must stay at the beginning and end position for a few seconds to start and stop the device, which would stack dozens of same frames in the data set. The hourglass network still output the predictions for all the stacked frames, and generated many invalid SP points at the beginning and end, as shown in white box in Fig. 9b and Fig. 10b. In some cases, the raters were disturbed by these scattered and invalid points when manually tuning the infection points of spinous process curve, and thus the measures of SPAs were affected, especially in the lumbar area.

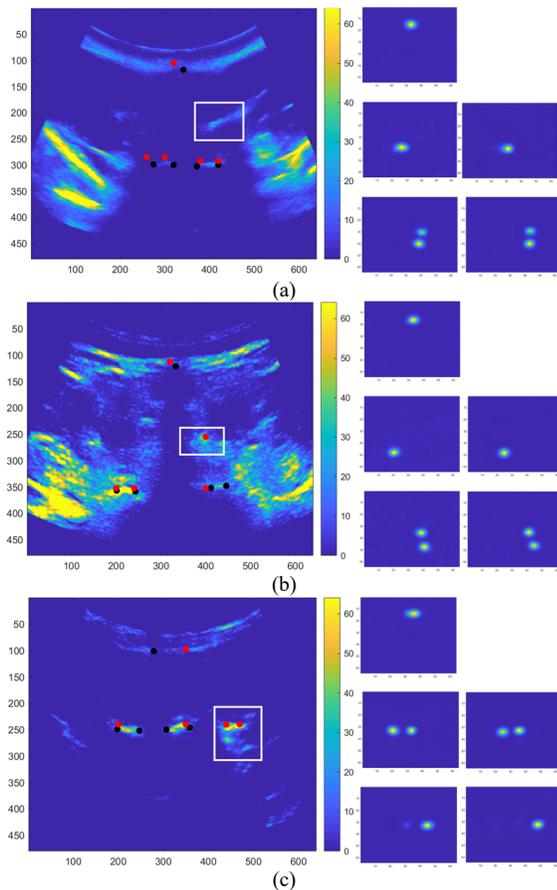

Fig. 8. Three types of cases resulting in bad predictions of SP position. The left column indicated the transverse images with red predicted landmarks and black ground truth. The right column indicated the 64*64 heatmaps corresponding to SP (1st row), LA0&LA1(2nd row), LA2&LA3(3rd row), respectively. The high intensity candidates in white box of (a), (b) were soft tissue and inferior articular process, which interfered the detection of right lamina landmarks LA2&LA3. The high intensity candidates in white box of (c) was rib, and it was detected as LA2 & LA3, therefore affected the detection of LA0&LA1.

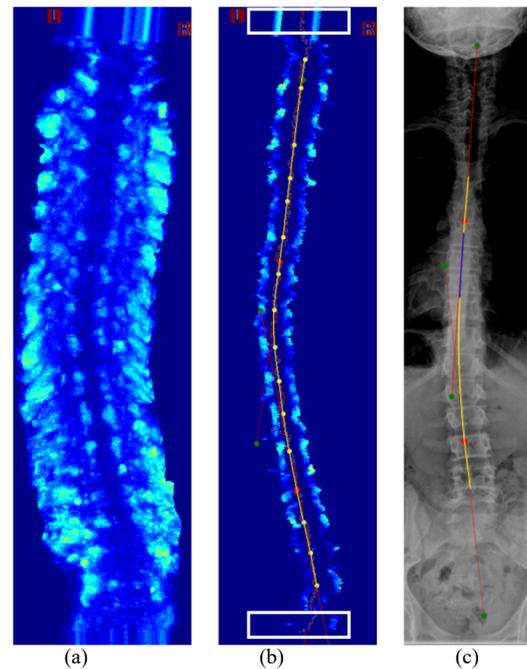

Fig. 9. The overestimate of SPA measurements: (a) original coronal image; (b) the segmented coronal image with SPA of 19°, and (c) radiograph with SPA of 11°. The white box illustrated the stacked frames as the beginning and end of US scan.

Ungi et al. [25] segmented the transverse process and ribs on 2-D transverse ultrasound images based on U-Net, and visualized them on coronal plane through 3-D reconstruction. The transverse process was proposed as an alternative way to measure scoliotic curves named as transverse process angle, which was the proxy of Cobb angle. Nevertheless, our study was the first study to detect spinous process and laminae as landmarks using deep learning methods on transverse ultrasound images. The SPA has been proved as a valuable clinical parameter to represent the posterior deformities on radiographs, and showed potential to predict the Cobb angle [31]. As show in Fig. 7a&b, the large differences and moderate correlation between ultrasound SPAs and radiographic Cobb angles indicated that there was inconsistency between these two evaluation parameters of scoliotic curves. It was mainly caused by the different landmarks used for the measurement, where the SPA used the spinous process and Cobb angle used the endplate of vertebra. As illustrated in the white box in Fig 5b&c, the SP curves on US image and radiograph were both deviated to the right direction, and the curvature was apparently different from the Cobb angle. The SPA usually was smaller compared to Cobb angle especially for large curvature, because the vertebral body generally rotate to the curve convexity and thus cause the SP tips lateral offset to the direction of the curve concavity [31]. However, the SPA is still a good evaluation parameter which can be simultaneously correlated with Cobb angle and apical vertebrae rotation (AVR).

The hourglass network presented higher efficiency and accuracy compared to the conventional active contour method such as GVF snake model [27]. Firstly, the GVF method required more computation time during the process. It took average 1.5 hours for one ultrasound scan to identify SP on each frame, while the average detection time for one scan using the stacked hourglass network was only 10 minutes. It implied that the hourglass network would be more suitable for clinical application and easier to deploy on portable devices. Secondly, the GVF method was easily influenced by the status of lamina pairs. It was required to firstly locate the laminae based on prior knowledge: the intensity of lamina is high. Therefore, the images with invisible lamina as show in Fig. 2 d-f would be skipped for SP detection when using the GVF method. Over all fifty US scans, the average percentage of valid detection was only 50% from the GVF method [27], but the number from SHN was 96.9%. More identified SP formed a more continuous spinous process curve on coronal plane and reduced the difficulty of the observation and curve-fitting during SPA measurement. Moreover, the SHN could utilized the accurate and numerous laminae to visualize and highlight the posterior part of spine.

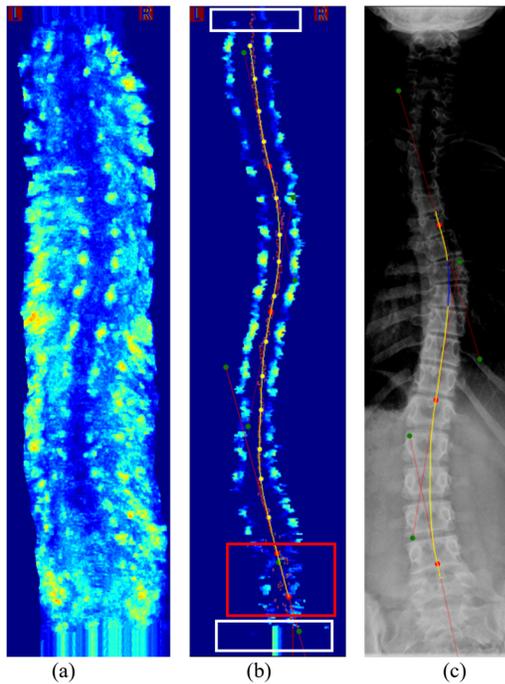

(a)    (b)    (c)

Fig. 10. The scattered predictions on lumbar region (red box), (a) the original coronal image. (b) the identified landmarks on the coronal plane, The SPAs were 24/26° on ultrasound coronal image; (c) the radiograph, and the SPAs were 26/18°. The white box illustrated the stacked frames as the beginning and end of US scan.

As mentioned in the previous section, the network was compelled to detect five landmarks on each US transverse image to improve the prediction results, therefore all the images in training dataset were selected to be located on vertebrae as displayed in Fig. 11 a & c. However, for entire US scan, there were a certain number of frames located at the intervals between two vertebrae. In these interval frames, the vertebral structures were not complete, for example the laminae were absent as illustrated in Fig .11b. For these frames, the present SHN still compulsively output the fake predictions of SP and laminae. Fortunately, the number of fake SP and lamina position was small comparing to the total number of frames in one data set, and it would not significantly deviate the tendency of spine curve and thus not influence the SPA measurement. However, avoiding fake detection was valuable for automatic measurement. When measuring the SPA, to reduce the variation and obtain a smooth curve, only a few featured points were selected to represent different vertebrae and then fit into the spine curve. The continuously distributed predicted SP from SHN made it impossible to automatically classify these points to different vertebral groups. To solve this problem, the frames in one US scan can be sorted into two categories according to whether locating on vertebra or gap, and only the frames on vertebra were input into SHN. The CNN can be used for the classification and the information of consecutive frames may be utilized.

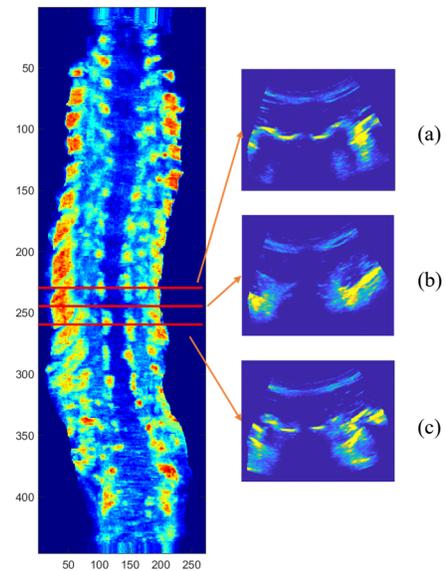

Fig. 11. The transverse frames in US scan. (a)&(c) located on vertebrae with clear laminae; (b) located on the interval between two vertebrae without laminae.

## V. CONCLUSION

In this study, a stacked hourglass network was trained and tested on ultrasound spine images to detect spinous process and lamina endpoints as five landmarks. A high average PCK of 86.8% was achieved for landmark detection, especially the PCK of SP detection was 90.3%. The SHN only spent 10 minutes averagely for processing one ultrasound scan and was appropriate for clinical application. The results showed a good correlation (>0.85) between US and radiographic measurements, and the MADs of SPAs from our ultrasound method and radiography were 3.3±2.6° and 3.3±2.3° for two raters, which were within the clinical acceptance error (5°). The proposed US method based on SHN can be regarded as a reliable and promising tool for diagnosis, monitoring and screening of spine deformities.

In the future work, a large scale of dataset with more complex vertebral structures will be implemented to improve the performance and generalization ability of stacked hourglass network for clinical application. On the other hand, new methods will be suggested to classify transverse images before detection, which could provide more efficient and accurate detection results, thus leading to better visualization of entire spine volume.

ACKNOWLEDGMENT

The authors would like to extend sincere gratitude to the Glenrose Hospital for providing the subjects' US scans in this study. And the authors also are profusely grateful to the sponsorship from Natural Science Foundation of Shanghai.

REFERENCES

[1] M. A. Asher and D. C. Burton, 'Adolescent idiopathic scoliosis: natural history and long term treatment effects', *Scoliosis*, vol. 1, no. 1, p. 2, Mar. 2006, doi: 10.1186/1748-7161-1-2.
[2] S. L. Weinstein, L. A. Dolan, J. C. Cheng, A. Danielsson, and J. A. Morcuende, 'Adolescent idiopathic scoliosis', *The Lancet*, vol. 371, no.

9623, pp. 1527–1537, May 2008, doi: 10.1016/S0140-6736(08)60658-3.
[3] H. Fan et al., 'Prevalence of Idiopathic Scoliosis in Chinese Schoolchildren', *SPINE*, vol. 41, pp. 259–264, Feb. 2016, doi: 10.1097/BRS.0000000000001197.
[4] J. COBB, 'Outline for the study of scoliosis', *Instr Course Lect AAOS*, vol. 5, pp. 261–275, 1948.
[5] H. Kim et al., 'Scoliosis Imaging: What Radiologists Should Know', *RadioGraphics*, vol. 30, no. 7, pp. 1823–1842, Nov. 2010, doi: 10.1148/rg.307105061.
[6] R. T. Morrissy, G. S. Goldsmith, E. C. Hall, D. Kehl, and G. H. Cowie, 'Measurement of the Cobb angle on radiographs of patients who have scoliosis. Evaluation of intrinsic error', *J. Bone Jt. Surg.*, vol. 72, no. 3, pp. 320–327, Mar. 1990, doi: 10.2106/00004623-199072030-00002.
[7] J. E. Herzenberg, N. A. Waanders, R. F. Closkey, A. B. Schultz, and R. N. Hensinger, 'Cobb Angle Versus Spinous Process Angle in Adolescent Idiopathic Scoliosis The Relationship of the Anterior and Posterior Deformities', *Spine*, vol. 15, no. 9, pp. 874–879, Sep. 1990, doi: 10.1097/00007632-199009000-00007.
[8] M.-H. Horng, C.-P. Kuok, M.-J. Fu, C.-J. Lin, and Y.-N. Sun, 'Cobb Angle Measurement of Spine from X-Ray Images Using Convolutional Neural Network', *Comput. Math. Methods Med.*, vol. 2019, pp. 1–18, Feb. 2019, doi: 10.1155/2019/6357171.
[9] H. Wu, C. Bailey, P. Rasoulinejad, and S. Li, 'Automatic Landmark Estimation for Adolescent Idiopathic Scoliosis Assessment Using BoostNet', in *Medical Image Computing and Computer Assisted Intervention − MICCAI 2017*, vol. 10433, M. Descoteaux, L. Maier-Hein, A. Franz, P. Jannin, D. L. Collins, and S. Duchesne, Eds. Cham: Springer International Publishing, 2017, pp. 127–135.
[10] H. Sun, X. Zhen, C. Bailey, P. Rasoulinejad, Y. Yin, and S. Li, 'Direct Estimation of Spinal Cobb Angles by Structured Multi-output Regression', in Information Processing in Medical Imaging, Cham, 2017, pp. 529–540. doi: 10.1007/978-3-319-59050-9_42.
[11] T. Zhang, Y. Li, J. P. Y. Cheung, S. Dokos, and K.-Y. K. Wong, 'Learning-based coronal spine alignment prediction using smartphone-acquired scoliosis radiograph images', *IEEE Access*, pp. 1–1, 2021, doi: 10.1109/ACCESS.2021.3061090.
[12] H. Wu, C. Bailey, P. Rasoulinejad, and S. Li, 'Automated comprehensive Adolescent Idiopathic Scoliosis assessment using MVC-Net', *Med. Image Anal.*, vol. 48, pp. 1–11, Aug. 2018, doi: 10.1016/j.media.2018.05.005.
[13] M. Law, W.-K. Ma, D. Lau, E. Chan, L. Yip, and W. Lam, 'Cumulative radiation exposure and associated cancer risk estimates for scoliosis patients: Impact of repetitive full spine radiography', *Eur. J. Radiol.*, vol. 85, no. 3, pp. 625–628, Mar. 2016, doi: 10.1016/j.ejrad.2015.12.032.
[14] A. K. Simpson, P. G. Whang, A. Jonisch, A. Haims, and J. N. Grauer, 'The Radiation Exposure Associated With Cervical and Lumbar Spine Radiographs', *Clin. Spine Surg.*, vol. 21, no. 6, pp. 409–412, Aug. 2008, doi: 10.1097/BSD.0b013e3181568656.
[15] C. M. Ronckers, C. E. Land, J. S. Miller, M. Stovall, J. E. Lonstein, and M. M. Doody, 'Cancer Mortality among Women Frequently Exposed to Radiographic Examinations for Spinal Disorders', *Radiat. Res.*, vol. 174, no. 1, pp. 83–90, Apr. 2010, doi: 10.1667/RR2022.1.
[16] M. Morin Doody et al., 'Breast Cancer Mortality After Diagnostic Radiography: Findings From the U.S. Scoliosis Cohort Study', *Spine*, vol. 25, no. 16, p. 2052, Aug. 2000.
[17] W. Chen, E. H. M. Lou, P. Q. Zhang, L. H. Le, and D. Hill, 'Reliability of assessing the coronal curvature of children with scoliosis by using ultrasound images', *J. Child. Orthop.*, vol. 7, no. 6, pp. 521–529, Dec. 2013, doi: 10.1007/s11832-013-0539-y.
[18] R. Zheng et al., 'Improvement on the Accuracy and Reliability of Ultrasound Coronal Curvature Measurement on Adolescent Idiopathic Scoliosis With the Aid of Previous Radiographs.', *Spine*, vol. 41, no. 5, pp. 404–411, 2016, doi: 10.1097/BRS.0000000000001244.
[19] C.-W. J. Cheung, G.-Q. Zhou, S.-Y. Law, T.-M. Mak, K.-L. Lai, and Y.-P. Zheng, 'Ultrasound Volume Projection Imaging for Assessment of Scoliosis', *IEEE Trans. Med. Imaging*, vol. 34, no. 8, pp. 1760–1768, Aug. 2015, doi: 10.1109/TMI.2015.2390233.
[20] R. Zheng et al., 'Intra- and Inter-rater Reliability of Coronal Curvature Measurement for Adolescent Idiopathic Scoliosis Using Ultrasonic Imaging Method—A Pilot Study', *Spine Deform.*, vol. 3, no. 2, pp. 151–158, Mar. 2015, doi: 10.1016/j.jspd.2014.08.008.
[21] R. Zheng et al., 'Factors influencing spinal curvature measurements on ultrasound images for children with adolescent idiopathic scoliosis (AIS)', *PLOS ONE*, vol. 13, no. 6, p. e0198792, Jun. 2018, doi: 10.1371/journal.pone.0198792.
[22] G.-Q. Zhou, W.-W. Jiang, K.-L. Lai, and Y.-P. Zheng, 'Automatic Measurement of Spine Curvature on 3-D Ultrasound Volume Projection Image With Phase Features', *IEEE Trans. Med. Imaging*, vol. 36, no. 6, pp. 1250–1262, Jun. 2017, doi: 10.1109/TMI.2017.2674681.
[23] H. Zeng, R. Zheng, L. H. Le, and D. Ta, 'Measuring Spinous Process Angle on Ultrasound Spine Images using the GVF Segmentation Method', in *2019 IEEE International Ultrasonics Symposium (IUS)*, Oct. 2019, pp. 1477–1480, doi: 10.1109/ULTSYM.2019.8925710.
[24] H.-Y. Zeng, E. Lou, S.-H. Ge, Z.-C. Liu, and R. Zheng, 'Automatic detection and measurement of Spinous Process Curve on Clinical Ultrasound Spine Images', *IEEE Trans. Ultrason. Ferroelectr. Freq. Control*, pp. 1–1, 2020, doi: 10.1109/TUFFC.2020.3047622.
[25] P. Wang, V. M. Patel, and I. Hacihaliloglu, 'Simultaneous Segmentation and Classification of Bone Surfaces from Ultrasound Using a Multi-feature Guided CNN', in Medical Image Computing and Computer Assisted Intervention – MICCAI 2018, Cham, 2018, pp. 134–142. doi: 10.1007/978-3-030-00937-3_16.
[26] A. Z. Alsinan, V. M. Patel, and I. Hacihaliloglu, 'Automatic segmentation of bone surfaces from ultrasound using a filter-layer-guided CNN', *Int. J. Comput. Assist. Radiol. Surg.*, vol. 14, no. 5, pp. 775–783, May 2019, doi: 10.1007/s11548-019-01934-0.
[27] T. Ungi et al., 'Automatic Spine Ultrasound Segmentation for Scoliosis Visualization and Measurement', *IEEE Trans. Biomed. Eng.*, vol. 67, no. 11, pp. 3234–3241, Nov. 2020, doi: 10.1109/TBME.2020.2980540.
[28] A. Newell, K. Yang, and J. Deng, 'Stacked Hourglass Networks for Human Pose Estimation', in Computer Vision – ECCV 2016, Cham, 2016, pp. 483–499. doi: 10.1007/978-3-319-46484-8_29.
[29] H. Chen, R. Zheng, E. Lou, and D. Ta, 'Imaging Spinal Curvatures of AIS Patients using 3D US Free-hand Fast Reconstruction Method', in *2019 IEEE* International *Ultrasonics Symposium (IUS)*, Oct. 2019, pp. 1440–1443, doi: 10.1109/ULTSYM.2019.8925758.
[30] F. Zhang, X. Zhu, H. Dai, M. Ye, and C. Zhu, 'Distribution-Aware Coordinate Representation for Human Pose Estimation', p. 10.
[31] D. G. Morrison, A. Chan, D. Hill, E. C. Parent, and E. H. M. Lou, 'Correlation between Cobb angle, spinous process angle (SPA) and apical vertebrae rotation (AVR) on posteroanterior radiographs in adolescent idiopathic scoliosis (AIS)', *Eur. Spine J.*, vol. 24, no. 2, pp. 306–312, Feb. 2015, doi: 10.1007/s00586-014-3684-1.